\documentclass[prd,a4paper,aps,showpacs,twocolumn,floats]{revtex4}
\usepackage{epsfig}
\usepackage{graphicx}
\usepackage{color}

\newcommand{\bfe}{{\mathbf e}}

\newcommand{\bn}{{\mathbf n}}

\newcommand{\De}{\Delta}
\newcommand{\ep}{\epsilon}

\newcommand{\si}{\sigma}

\newcommand{\be}{\begin{equation}}
\newcommand{\ee}{\end{equation}}

\newcommand{\bea}{\begin{eqnarray}}
\newcommand{\eea}{\end{eqnarray}}
\newcommand{\bean}{\begin{eqnarray*}}
\newcommand{\eean}{\end{eqnarray*}}

\begin{document}

\title{The local B-polarization of the CMB: a very sensitive probe of cosmic defects} 

\author{Juan  Garc\'{\i}a-Bellido$^{1,2}$, Ruth Durrer$^{2}$, Elisa Fenu$^{2}$, 
Daniel G. Figueroa$^{1,3}$, Martin Kunz$^{2}$\\[-2mm]  $÷$}

\affiliation{${}^{1}$\,Instituto de F\'{\i}sica Te\'orica CSIC-UAM and Departamento de 
F\'isica Te\'orica, Universidad Aut\'onoma de Madrid, Cantoblanco 28049 Madrid, Spain\\
${}^{2}$\,D\'epartement de Physique Th\'eorique, Universit\'e de Gen\`eve, 24 quai Ernest 
Ansermet, CH--1211 Gen\`eve 4, Switzerland\\
${}^{3}$\,Theory Division CERN, CH-1211 Gen\`eve 23, Switzerland}

\preprint{IFT-UAM/CSIC-10-10, CERN-PH-TH/2010-054}

\date{November 11, 2010}

\begin{abstract}
We present a new and especially powerful signature of 
cosmic strings and other topological or non-topological defects in the 
polarization of the cosmic microwave background (CMB). We show that 
even if defects contribute 1\% or less in the CMB temperature anisotropy 
spectrum, their signature in the local $\tilde{B}$-polarization 
correlation function at angular scales of tens of arc-minutes is much larger 
than that due to gravitational waves from inflation, even if the latter 
contribute with a ratio as big as $r\simeq 0.1$ to the temperature 
anisotropies. We show that when going from non-local to local 
$\tilde{B}$-polarization, the ratio of the defect signal-to-noise with respect 
to the inflationary value increases by about an order of magnitude.
Proposed B-polarization experiments, with a good sensitivity on arc-minute 
scales, may either detect a contribution from topological defects
produced after inflation or place stringent limits on them. Already Planck 
should be able to improve present constraints on defect models by about 
an order of magnitude, to the level of $\ep=Gv^2 <10^{-7}$. A future 
full-sky experiment like CMBpol, with polarization sensitivities of the order 
of $1\mu$K-arcmin, will be able to constrain the defect parameter $\ep$ 
to less than a few $\times10^{-9}$, depending on the defect model. 
\end{abstract}


\maketitle

\section{Introduction}

Many inflationary models terminate with a phase transition which often
also leads to the formation of cosmic strings and other topological 
defects~\cite{mairi}.
Furthermore, we have recently argued~\cite{our} that the end of
hybrid inflation may involve the self-ordering of a $N$-component
scalar field. Even though for $N>4$ it does not lead to the formation of
topological defects, the self-ordering dynamics leads to a scale-invariant 
spectrum of fluctuations which leaves a signature on the 
CMB~\cite{largeN,texture}. It has been shown long ago that
topological defects do not generate acoustic peaks~\cite{dsg} and therefore
they cannot provide the main contribution to the CMB anisotropies. However,
they still may provide a fraction of about 10\%, similar to a possible 
gravitational wave contribution~\cite{defCMB} in the temperature anisotropies 
of the CMB.

The perturbations from cosmic strings and other topological defects
are proportional to the dimensionless variable $\ep=Gv^2$ where $v$ is 
the symmetry breaking scale. For cosmic strings $\mu=v^2$ is the 
energy per unit length of the string~\cite{defrev}. Present CMB
data limit the contribution from defects~\cite{defCMB} 
such that $\ep < 7\times 10^{-7}$. 
Stronger limits on $\ep$ have been derived from the gravitational waves
emitted from cosmic string loops~\cite{VV,D,pulsar}, but these are quite 
model dependent and will not be discussed here.

In this Letter we show that measuring the local $\tilde{B}$-polarization 
correlation function of the CMB provides stringent limits
on defects or, alternatively, detects them. 
The physical reason for this is twofold. First,
defects lead not only to tensor but also to even larger vector 
perturbations~\cite{texture}. What is more important, vector modes generate
much stronger B-polarization than tensor modes with the same amplitude, see 
e.g.~\cite{mybook}. B-polarization is not only a `smoking gun' for 
gravitational waves from inflation, but it is also extremely sensitive to the 
presence of vector perturbations (vorticity).
Furthermore, the B-polarization of the angular power spectrum of topological 
defects, especially of cosmic strings, peaks on somewhat smaller scales than 
the one from tensors due to inflation. The local $\tilde B$-correlation 
function, which is obtained from the polarization by two additional 
derivatives, enhances fluctuations on small angular scales. As we shall see, 
measuring the local $\tilde{B}$ instead of the usual non-local $B$ correlation
function results in an enhancement of the signal to noise ratio
from defects with respect to the inflationary one by about a factor 10.

\section{The local $\tilde B$-polarization correlation function}

Since Thomson scattering is direction dependent, a non-vanishing quadrupole 
anisotropy on the surface of last scattering leads to a slight polarization 
of the CMB~\cite{mybook}. This polarization is described as 
a rank-2 tensor field ${\cal P}_{ab}$ on the sphere, the CMB sky.
It is usually decomposed into Stokes parameters,
${\cal P}_{ab}=(I\si_{ab}^{(0)} +U\si_{ab}^{(1)} +V\si_{ab}^{(2)}+
Q\si_{ab}^{(3)})/2=I\delta_{ab}/2 +P_{ab}$,
where $\si^{(\mu)}$ are the Pauli matrices~\cite{mybook},
and $I$ corresponds to the intensity of the radiation and contains the 
temperature anisotropies. Thomson scattering
does not induce circular polarization so we expect $V=0$ for the CMB
polarization, and hence $P_{ab}$ to be real.
We define an orthonormal frame 
$(\bfe_1,\bfe_2,\bn)$ and the circular polarization vectors 
$\bfe_\pm = \frac{1}{\sqrt{2}}\left(\bfe_1\pm i\bfe_2\right)$, which allows us
to introduce the components $P_{\pm\pm}= 2\bfe_\pm^a \bfe_\pm^bP_{ab} 
= Q \pm i U$ and $P_{+-} \sim V =0$.  The second derivatives
of this polarization tensor are related to the local $\tilde{E}$- and 
$\tilde{B}$-polarizations,
\bea\nonumber
\nabla_-\!\nabla_-P_{++} + \nabla_+\!\nabla_+P_{--} &\!=\!& 2\nabla_a\!\nabla_bP_{ab}
 \equiv \tilde E\,, \\ \nonumber
\nabla_-\!\nabla_-P_{++} - \nabla_+\!\nabla_+P_{--} &\!=\!& 2\ep_{cd}\ep_{ab}
\nabla_c\!\nabla_aP_{bd} \equiv \tilde B \,. \hspace{2mm}
\eea
Here $ \nabla_{\pm}$  are the derivatives in the directions 
$ \bfe_{\pm} $ and
$\ep_{cd}$ is the 2-dimensional totally anti-symmetric tensor.
These functions are defined {\em locally}. The usual $E$- and $B$-modes 
can be obtained by applying the inverse Laplacian to the local $\tilde E$- and 
$\tilde B$-polarizations. Such inversions of differential operators depend on boundary 
conditions which can affect the result for local observations.
The $\tilde{B}$-correlation function, $C^{\tilde B}(\theta) \equiv  \langle 
\tilde B(\bn)\tilde B(\bn')\rangle_{\bn\cdot\bn' =\cos\theta}$,
is measurable locally. It is related to the $B$-polarization power spectrum
$C_{\ell}^{B}$ by~\cite{mybook}
\be\label{CBL}
C^{\tilde B}(\theta) =\frac{1}{4\pi}\sum_{\ell=2}^\infty 
 \frac{(\ell+2)!}{(\ell-2)!}(2\ell+1)P_\ell(\cos\theta)
C_{\ell}^{B} \,.
\ee
Here $P_\ell(x)$ are the Legendre polynomials. Analogous formulae also 
hold for $C^{\tilde E}$. Note the additional factor
$n_\ell = (\ell+2)!/(\ell-2)!=\ell(\ell^2-1)(\ell+2)\sim \ell^4$ as compared 
to the usual non-local $E$- and $B$-polarization correlation functions. 
At first sight one might argue that whether one expresses
a result in terms of $C_{\ell}^{B}$'s or $C_{\ell}^{\tilde B}=n_\ell C_{\ell}^{B}$ 
should really not make a difference since both contain the same information. 
For an ideal full sky experiment which directly measures the $C_{\ell}^{B}$ 
with only instrumental errors this is true. But a CMB experiment usually 
measures a polarization direction and amplitude with a given resolution 
over a patch of sky and with a significant noise level, and this makes a big 
difference as we shall show.

\section{Results}

\begin{figure}[th]
\centerline{\includegraphics[width=8.1cm]{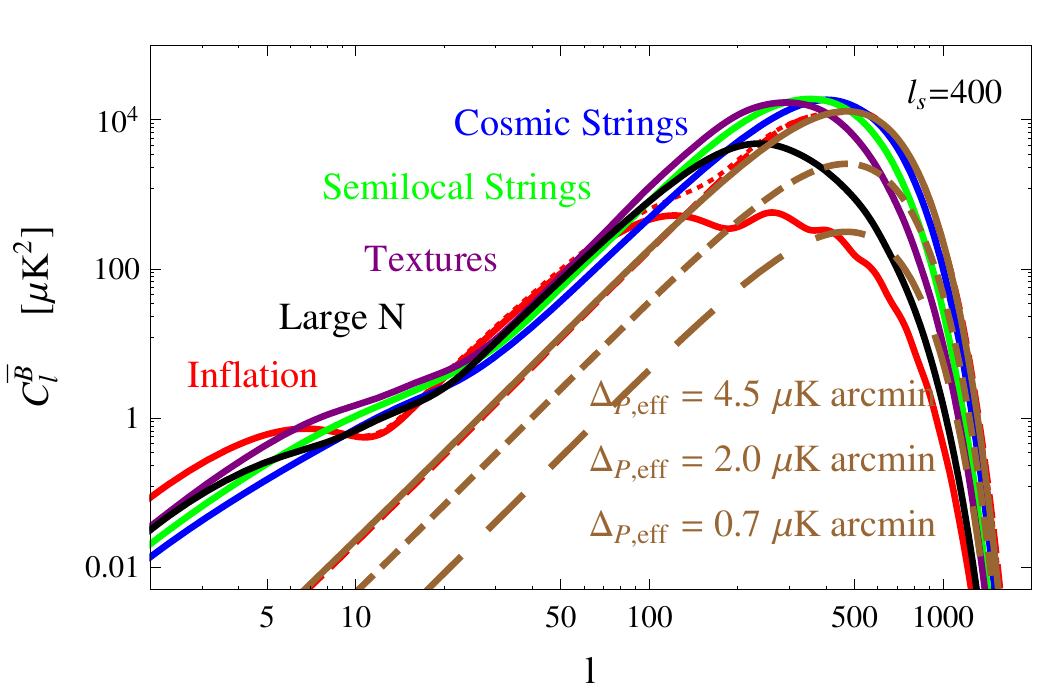}}
\caption{ \label{f:CBl} 
The local $\tilde{B}$-polarization power spectra for tensor perturbations
from inflation, cosmic strings, textures and the large-$N$ limit of the 
non-linear sigma-model. All spectra are normalized such that they make up 
10\% of the temperature anisotropy at $\ell=10$. The dotted red line corresponds
to the inflationary contribution taking into account the induced power 
from lensing of E-modes. The different noise levels (dashed brown curves) 
precisely mimic the effect of E-lensing. For a definition of the noise amplitude 
$\De_{P,\rm eff}$ and the smoothing scale $\ell_s$ see the text.
}
\end{figure}

In Fig.~\ref{f:CBl} we show the local $\tilde B$-polarization power spectra 
for tensor perturbations from inflation,
cosmic strings, textures and the large-$N$ 
limit of the non-linear sigma-model. All spectra are normalized such that they 
make up 10\% of the temperature anisotropy at $\ell=10$. Details of how these 
calculations are done can be found in~\cite{texture} for 
global defects and the large-$N$ limit and in~\cite{meth} for cosmic strings. 
A comparison of the non-local $B$-polarization power spectra for cosmic 
strings and inflation can be found in~\cite{comp1}.

It had already been noted in Refs.~\cite{SPT} and~\cite{defrev} that the 
$B$-polarization power spectra for defects are larger than those from inflation
for the same temperature anisotropy. Defects peak at somewhat higher $\ell$'s  
than inflationary perturbations, since  $B$-modes from 
defects are dominated by their vector (vorticity) modes. This  contribution
is maximal on scales that are somewhat smaller than the horizon scale, while
gravitational waves truly peak at the Hubble horizon at decoupling, 
which corresponds to $\ell \sim 100$.  As a consequence, the {\em local} 
$\tilde B$-polarization spectra for defects are even larger than  those from 
inflation because of the factor $n_\ell\simeq \ell^4$.
This is most pronounced for cosmic
strings, which have considerable power on small scales, but it is also true 
for other defects.

\begin{figure}[th]
\centerline{\includegraphics[width=8.6cm]{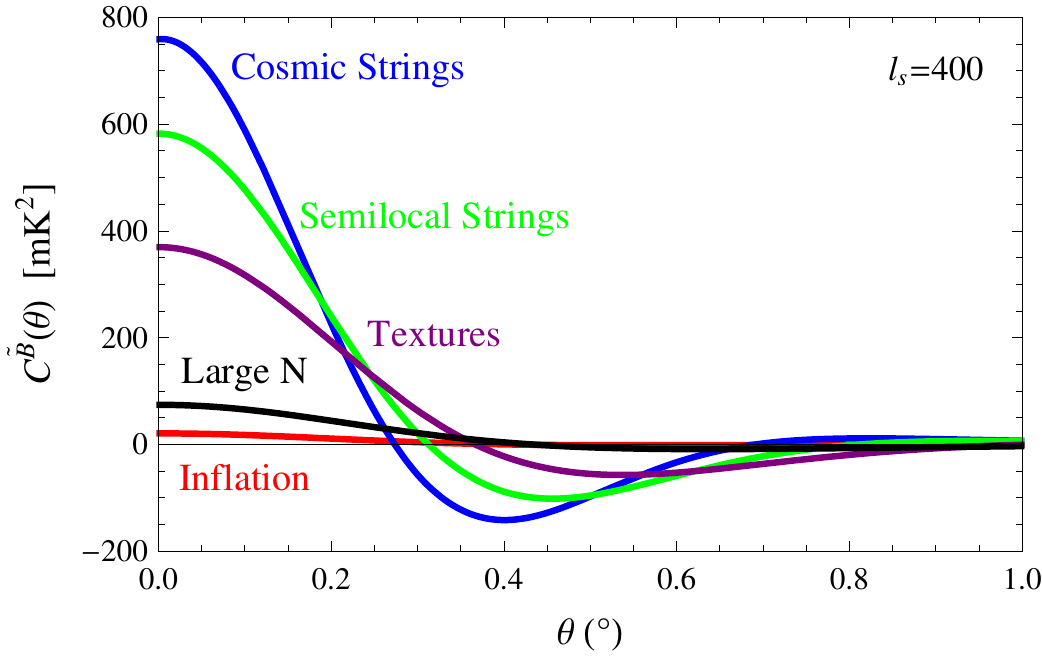}}
\caption{ \label{f:CBth} 
The local $\tilde{B}$-polarization angular correlation functions for $\theta<1^o$ for
inflation and the defect models of Fig.~1, with a smoothing scale $l_s=400$. 
}
\end{figure}

Due to the extra factor $n_\ell$, in the local $\tilde B$-power spectra shown in 
Fig.~\ref{f:CBl}, power at small scales (high $\ell$) counts 
significantly more than power at larger scales (low $\ell$). This is the reason why 
defect models dominate over the inflationary
$B$-modes of the same amplitude. This is seen very prominently in the 2-point
angular correlation function shown in  Fig.~\ref{f:CBth}
where we can compare the defect peaks coming from cosmic strings, 
textures and large-$N$.
Note the decreasing height but increasing width of the peak as we go from
cosmic strings to large-N models.

For $0.2<\theta <1^o$, where the inflationary $\tilde B$-polarization
is about $-2$ mK$^2$, that from cosmic strings is 
$-150$ mK$^2$, about a factor 100 larger. For textures and the large-N model,
the difference is somewhat smaller, roughly  a factor of 50 and 10 respectively. 
The very pronounced peak on very small scales is not visible due to the noise.

Even though constructed ad hoc, 
coherent causal seed models (but not topological defects) can 
have acoustic peaks, see Ref.~\cite{Tukmod}, which thus cannot be used as a 
differentiating signature from inflation. But the fact that polarization 
is generated at the last scattering surface implies that it cannot have power 
on scales larger than the horizon at decoupling, corresponding to about 
$\ell\sim 100$, or angles $\theta > 2^o$, unless something like inflation has 
taken place~\cite{SZ}. This can only be circumvented if one allows for 
acausality, i.e. superluminal motion, of the seeds~\cite{acausal}, however 
improbable. In Ref.~\cite{BZ} the authors have shown that this superhorizon 
signature appears not only in the TE-cross correlation spectrum, but also in
the local $\tilde B$-polarization spectrum. We find that this is
somewhat weakened by re-ionization, which adds power on large 
scales to the $B$-polarization from defects, see Fig.~\ref{f:CBl}.

\section{Observational prospects}

It is clear from Fig.~\ref{f:CBth} that cosmic defects with equal amplitude 
as the tensor component from inflation (note $\ep=7\times10^{-7}$ is 
equivalent to $r=0.1$) would have a significant peak in the two-point 
correlation function of the local $\tilde B$-polarization, on angular scales of
order tens of arc-minutes. A relevant issue is whether this peak could be 
measured with full-sky probes like Planck~\cite{planck} or 
CMBpol~\cite{CMBpol}, or even with small-area experiments. 
This is difficult because, although CMB experiments typically have a flat 
(white) noise power spectrum for the Stokes parameters, the 
{\em local} $n_\ell\sim \ell^4$ factor induces a very blue spectrum for the 
noise in the local $\tilde B$-modes, which erases the significance of the 
broad defect peak at $\ell\sim500$ in the $C_\ell^{\tilde B}$ power spectrum. 
Moreover, in order to extract the cosmological $\tilde B$-polarization signal 
it is necessary first to clean the map from the contribution coming from 
gravitationally lensed $\tilde E$-modes. This induces an extra `lensing noise' 
$\Delta_{P,{\rm eff}} \sim 4.5\,\mu$K$\cdot$arcmin for
uncleaned maps that can be reduced to $\sim (0.1 - 0.7)\,\mu$K$\cdot$arcmin 
by iterative cleaning or a simple quadratic estimator 
respectively~\cite{SelHir}. Furthermore, 
CMB experiments have an angular resolution determined by the microwave horn 
beam width, $\theta_{\rm FWHM}$, which induces an uncertainty in the $C_\ell$'s 
that can be described by an exponential factor $\exp[\ell(\ell+1)\sigma_b^2]$, 
with $\sigma_b = \theta_{\rm FWHM}/\sqrt{8\log 2}$. Resolutions of order 10 
arc\-minutes, like those of the Planck HFI experiment, correspond to multipoles 
$\ell_b=1/\sigma_b\sim 800$. Adding the steep polarization noise, with typical 
amplitude $\Delta_{P,{\rm eff}} = (0.5 - 12)\,\mu$K$\cdot$arcmin, would make 
the signal disappear under the small-scale noise. 
In order to regulate this divergence, we smooth both the signal and the noise 
with a Gaussian smoothing of width $\sigma_s$, corresponding to a smoothing 
scale $\ell_s < \ell_b$. We choose $\ell_s=400$ in our analysis.

\begin{figure}[th]
\centerline{\includegraphics[width=8.1cm]{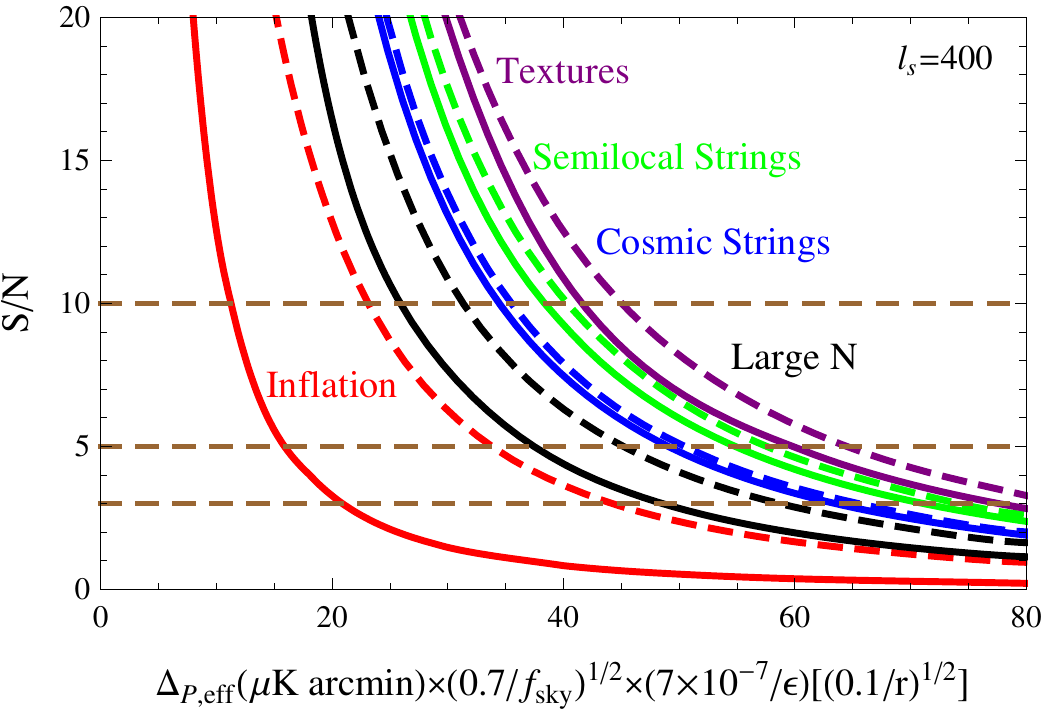}}
\caption{ \label{f:SNR} 
The signal-to-noise ratio as a function of the normalized polarization 
sensitivity, for inflation, cosmic strings, textures and the large-$N$ limit 
of the non-linear sigma-model. 
Solid curves: using angular scales up to $1^o$ and dashed curves: using 
angular scales up to $4^o$, with 6 arcmin resolution bins.}
\end{figure}

In order to compute the signal-to-noise ratio $S/N$ for detection of the defect 
peak in the local $\tilde B$-correlation function, we split the interval 
$\theta\in[0,1^o]$ in 10 equal bins~\footnote{Note that 
Planck has this resolution only for the higher frequency bands, above 200 GHz, 
where the sensitivity is somewhat reduced.}.  We then evaluate the theoretical 
correlation function at the center of those bins, 
${\sf S}_i = C^{\tilde B}(\theta_i)$, and write 
the covariance matrix of the correlated bins as
$${\sf C}_{ij} = \sum_\ell \frac{2\ell+1}{8\pi^2f_{\rm sky}} 
({\cal C}_\ell^{\tilde B})^2P_\ell(\cos\theta_i)P_\ell(\cos\theta_j)\,,$$
where the covariance matrix in $\ell$-space is assumed to be diagonal, 
${\rm cov}[{\cal C}_\ell^{\tilde B},{\cal C}_{\ell'}^{\tilde B}]=
2({\cal C}_\ell^{\tilde B})^2\delta_{\ell\ell'}/(2\ell+1)f_{\rm sky}$, 
with ${\cal C}_\ell^{\tilde B} = (C_\ell^{\tilde B} + N_\ell)
\exp[-\ell(\ell+1)/\ell_s^2]$. Here $f_{\rm sky}$ is the fraction of the 
observed sky which we set to 0.7 for satellite probes.
The signal-to-noise ratio for the defect model is 
$S/N = \sqrt{{\sf S}_i {\sf C}^{-1}_{ij} {\sf S}_j  }$. In Fig.~\ref{f:SNR} 
we show this ratio as a function of the normalized polarization sensitivity 
for all types of defects as well as for inflation (where $7\times 10^{-7}/\ep$ 
has to be replaced by $\sqrt{0.1/r}$). The horizontal
lines correspond to 3, 5 and 10-$\sigma$ respectively. To show why the choice
of $\theta_{\max}=1^o$ is optimal we also plot (dashed lines) the $S/N$ for 
$\theta_{\max}= 4^o$, at fixed resolution ($6'$). For the latter, the noise 
level allowed for a 3-$\sigma$ detection increases by more 
than a factor of 2 for inflation while it does not change much for defects. 
This behaviour is a telltale sign for defects, and shows that their
signal is strongly localised in the angular correlation function,
which distinguishes them e.g. from inflationary tensor perturbations and lensed 
E-modes: the $S/N$ curve from defects does not change much for 
angles above $\sim1^o$, while the one from inflation increases significantly.

 In Table~\ref{tab:3sigma} we give the values of $\ep$ which are measured at 
3$\sigma$ by Planck (assuming $\Delta_{P,{\rm eff}} = 
11.2\,\mu$K$\cdot$arcmin~\cite{BZ}, where the de-lensing error is added in 
quadrature), a CMBpol-like experiment with polarization sensitivity 
$\Delta_{P,{\rm eff}}=0.7\,\mu$K$\cdot$arcmin, and a dedicated CMB experiment 
with $\Delta_{P,{\rm eff}}=0.01\,\mu$K$\cdot$arcmin. 
Note, however, that it is not clear how to perform the de-lensing of the 
$B$-modes to the level of precision needed for the last case.

\begin{figure}[th]
\centerline{\includegraphics[width=8.1cm]{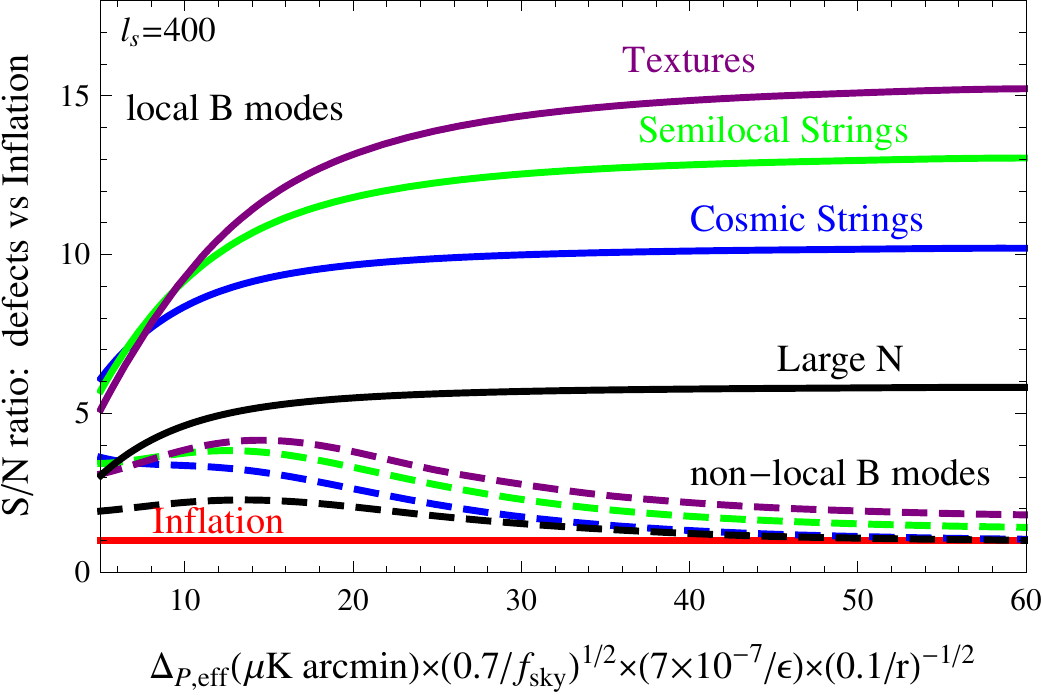}}
\caption{ \label{f:SNRr} 
The ratio of the signal-to-noise from defects to the one from
inflation. Solid curves: measuring the local $\tilde B$ correlation
function. Dashed curves: measuring the non-local $B$ correlation
function.  
}
\end{figure}

In Fig.~\ref{f:SNRr} we show the ratio of $S/N$ from defects to 
the one from inflation for non-local (dashed) and local $\tilde B$-modes
(solid curves). Clearly, in the local polarization the defect signal
is substantially enhanced.
It is interesting to note that actually textures fare better than cosmic 
strings even though they have less power on small scales. The reason is 
that the very small scales are dominated by noise and the signal mainly comes 
from the intermediate scales around $0.3^o$ where textures dominate, see Fig.~2.

\begin{table}
\caption{The limiting amplitude, $\ep=Gv^2$, of various defects,
at 3-$\sigma$ in the range $\theta\in[0,1^o]$, for Planck ($\Delta_{P,{\rm eff}}=
11.2\,\mu$K$\cdot$arcmin), CMBpol-like exp. 
($\Delta_{P,{\rm eff}}=0.7\,\mu$K$\cdot$arcmin) 
and a dedicated CMB experiment with ($\Delta_{P,{\rm eff}}=0.01\,
\mu$K$\cdot$arcmin). We set $f_{\rm sky}=0.7$.
\label{tab:3sigma}}
\vspace*{5mm}
\centering
\begin{tabular}{|c|cccc|}
\hline
$S/N=3$ & Strings & Semi-local & Textures & Large-N \\
\hline
Planck & $1.2\cdot10^{-7}$ & $1.1\cdot10^{-7}$ & $1.0\cdot10^{-7}$ & $1.6\cdot10^{-7}$  \\
\hline
CMBpol & $7.7\cdot10^{-9}$ & $6.9\cdot10^{-9}$ & $6.3\cdot10^{-9}$ & $1.0\cdot10^{-8}$  \\
\hline
$\tilde B$ exp & $1.1\cdot10^{-10}$ & $1.0\cdot10^{-10}$ & $0.9\cdot10^{-10}$ & $1.4\cdot10^{-10}$  \\
\hline
\end{tabular}
\end{table}

\section{Conclusions}

In this Letter we have shown that measuring  the local 
$\tilde B$-polarization correlation function on small scales, $\theta\lesssim 1^o$
is a superb way to detect topological and non-topological defects, or 
alternatively to constrain their contribution to the CMB. For simple 
inflationary models which lead to defect formation at the end of inflation, 
a value of $\ep\simeq 10^{-7}\, \div \, 10^{-8}$
seems rather natural, hence the achieved limits include the relevant regime. 
The fact that the local $\tilde B$-polarization 
from defects is dominated by the vector mode, which peaks on scales smaller 
than the horizon, is responsible for a significant enhancement of the local 
$\tilde B$-polarization correlation function on tens of arc-minute scales.

Even though the Planck satellite is not the ideal probe for constraining these 
models, if it finally reaches down to $r\le 0.025$, see Ref.~\cite{ESAdoc}, it
will either lead to the detection of a defect contribution, or it will constrain it 
to $\ep = Gv^2 \lesssim 10^{-7}$, depending on the defect model 
(textures being the 
most constrained and Large-N non-topological defects the least). Future 
CMB experiments, with 0.1 arc-minute resolution and sensitivities
at the level of $0.1\,\mu$K in polarization, could in principle reach the bound
$\ep < 10^{-10}$ for most defect types, which would rule out a large fraction 
of present models.

\section*{Acknowledgements}

We thank Neil Bevis, Mark Hindmarsh and Jon Urrestilla for
allowing us to use their $C_\ell$'s from cosmic string and texture simulations.
D.G.F. acknowledges support from a Marie Curie Early
Stage Research Training Fellowship associated with the EU RTN
ÒUniverseNetÓ during his stay at CERN TH-Division. J.G.B. thanks
the Institute de Physique Theorique de lÕUniversite de Geneve for
their generous hospitality during his sabbatical in Geneva. This
work is supported by the Spanish MICINN under project 
AYA2009-13936-C06-06 and by the EU FP6 Marie Curie Research 
and Training Network ÒUniverseNetÓ (MRTN-CT-2006-035863).


\end{document}